\documentclass[10pt,%
 aps,
groupedaddress,
 amsmath,amssymb,
prl,
twocolumn,superscriptaddress,
floatfix,
]{revtex4-2}

\usepackage{graphicx}
\usepackage{dcolumn}
\usepackage{bm}
\usepackage{relsize}
\usepackage{siunitx}
\usepackage{nicefrac}
\DeclareSIUnit\bar{bar}
\usepackage{graphics}
\usepackage{tabularx}
\usepackage{ulem}
\usepackage{xcolor}

\newcommand{\hci}[3]{\textsuperscript{#1}#2\textsuperscript{#3}} 
\newcommand{\alp}[0]{A{\smaller \smaller LPHATRAP}}

\newcommand{\Center}[1]{\multicolumn{1}{c}{#1}}
\newcommand{\Ion}[3]{${}^{#1}\mathrm{#2}^{#3}$}
\newcommand{\Neg}[1]{-#1}
\newcommand{\Pos}[1]{\phantom{-}#1}

\begin{document}

\preprint{APS}

\title{
\textit{g} Factor of Boron-like Tin}
\author{J.~Morgner\textsuperscript{$\dagger$}}
\email{jonathan.morgner@mpi-hd.mpg.de \textsuperscript{$\dagger$}These authors contributed equally}
\affiliation{Max-Planck-Institut für Kernphysik, Heidelberg, Germany}
\author{B.~Tu\textsuperscript{$\dagger$}}
\affiliation{Max-Planck-Institut für Kernphysik, Heidelberg, Germany}
\affiliation{Shanghai EBIT laboratory, Institute of Modern Physics, Fudan University, Shanghai, China}
\author{M.~Moretti}
\affiliation{Max-Planck-Institut für Kernphysik, Heidelberg, Germany}
\author{C.~M.~König}
\affiliation{Max-Planck-Institut für Kernphysik, Heidelberg, Germany}
\author{F.~Heiße}
\affiliation{Max-Planck-Institut für Kernphysik, Heidelberg, Germany}
\author{T.~Sailer}
\affiliation{Max-Planck-Institut für Kernphysik, Heidelberg, Germany}
\author{V.~A.~Yerokhin}
\affiliation{Max-Planck-Institut für Kernphysik, Heidelberg, Germany}
\author{B.~Sikora}
\affiliation{Max-Planck-Institut für Kernphysik, Heidelberg, Germany}
\author{N.~S.~Oreshkina}
\affiliation{Max-Planck-Institut für Kernphysik, Heidelberg, Germany}
\author{Z.~Harman}
\affiliation{Max-Planck-Institut für Kernphysik, Heidelberg, Germany}
\author{C.~H.~Keitel}
\affiliation{Max-Planck-Institut für Kernphysik, Heidelberg, Germany}
\author{S.~Sturm}
\affiliation{Max-Planck-Institut für Kernphysik, Heidelberg, Germany}
\author{K.~Blaum}
\affiliation{Max-Planck-Institut für Kernphysik, Heidelberg, Germany}

\date{\today}

\begin{abstract}
In the \alp~experiment, the $g$ factor of boron-like \hci{118}{Sn}{45+} has been measured with a $0.5$ parts-per-billion uncertainty.
This is the first high-precision measurement of a heavy boron-like $g$ factor.
The measured value of $0.644\,703\,826\,5(4)$ is consistent with the presented \textit{ab initio} state-of-the-art theory calculations, which predict a value of $0.644\,702\,9(8)$.
So far, the only boron-like $g$ factor measured with high precision has been \hci{40}{Ar}{13+}.
The measurement presented here therefore tests quantum electrodynamics as well as many-electron interactions at much higher $Z$.
Furthermore, we discuss the potential for an independent determination of the fine-structure constant $\alpha$, which can be achieved with a specific difference of $g$ factors, combining the presented results with the recent electron $g$-factor measurement of hydrogen-like tin.

\end{abstract}

\maketitle
Highly charged ions (HCI), which make up the majority of visible matter in the universe, are of great interest in astrophysical and plasma physics research.
With only a few bound electrons and a substantial nuclear charge $Z$, they offer unique opportunities to test fundamental theories~\cite{jentschura_lamb_1997,mohr_qed_1998,beier_g_j_2000,kozlov_highly_2018}.
For many decades, numerous experiments were performed to confirm the validity of quantum electrodynamics (QED) using HCI, both with hydrogen-like ions \cite{gumberidze_quantum_2005,ullmann_high_2017, morgner_stringent_2023} and in multi-electron systems~\cite{gould_radiative_1974,livingston_fine_1995,beier_determination_2003, beiersdorfer_measurement_2005,ullmann_high_2017,arapoglou_g_2019,silwal_spectroscopic_2021,loetzsch_testing_2024}.

As a cornerstone of the Standard Model, QED describes the interactions between charged particles with the inclusion of quantum effects such as self energy (SE) and vacuum polarisation (VP).
The magnetic dipole moment of an electron, typically represented by the $g$ factor, can be calculated precisely within the framework of QED.
Consequently, it serves as an ideal candidate for a rigorous QED assessment by comparing precise theoretical and experimental values.

\begin{figure*}[t]
    \centering
    \includegraphics{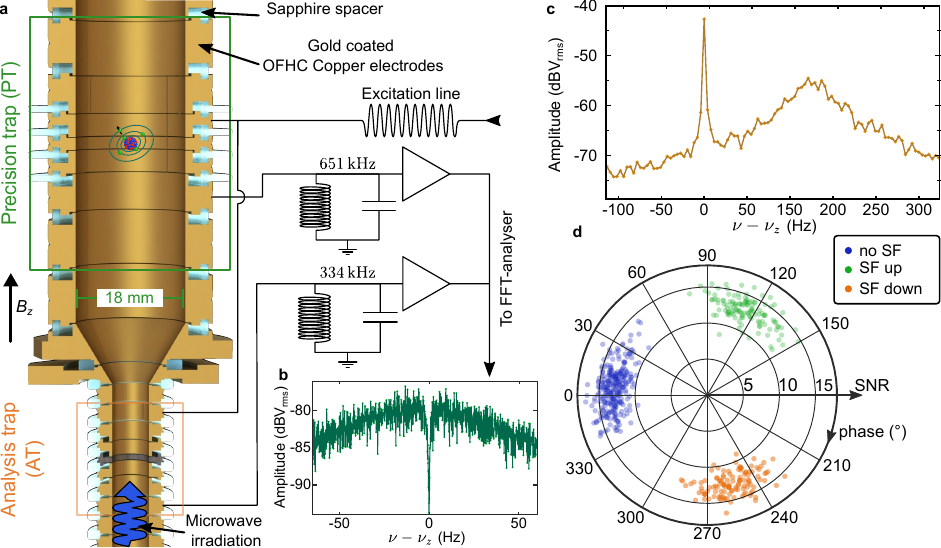}
    \caption{
    \textbf{a} shows a cut view of the double-trap setup, consisting of the large PT for the precise measurement of $\Gamma_0$ and the AT incorporating an iron-cobalt ring electrode (marked in black) used to determine the spin orientation of the ion.
    \textbf{b} shows the \textit{dip} signal of a particle in thermal equilibrium with the detector system in the PT where the axial frequency is determined.
    The spin orientation is determined by precise frequency measurements: The axial amplitude is excited by a resonant radio-frequency pulse.
    After a free evolution time the phase of the axial motion is determined from the peak in the FFT spectrum spectrum shown in~\textbf{c}.
    Irradiating a microwave at the Larmor frequency can \textit{flip} the spin. The evolution time is set such that a spin flip corresponds to a 115\,$^\circ$ degree phase shift of the axial motion.
    Whether the phase changes upwards or downwards depends on the direction of the spin change $\Delta m_j = \pm 1$ and therefore includes the information of the spin state.
    \textbf{d} shows the phase data of the AT spin flip (SF) detection accumulated over the $g$-factor measurement, demonstrating excellent readout of the 105\,mHz frequency change, and hence of the spin state.
    The three cluster represent the three cases where the microwave did not flip the spin (blue), those where the spin change was $\Delta m_j = 1$ (green) and those with $\Delta m_j = -1$ (orange).
    }
    \label{fig:trap}
\end{figure*}

For electrons in the $s$ state, the $g$ factors are close to that of the free electron, whereas for the $l>0$ states it is significantly different due to the presence of orbital angular momentum.
Additionally, in multi-electron systems, the interaction between bound electrons has to be taken into account for a precise $g$-factor calculation.
Thus, measurements of lithium-like and boron-like charge states are optimal for testing this contribution.
Until now, the only high-precision boron-like $g$-factor measurement has been performed with \hci{40}{Ar}{13+} ($Z=18$), reaching an experimental accuracy of 1.4~parts-per-billion~\cite{arapoglou_g_2019}.
The $g$ factor of its first excited state has been measured as well, reaching a precision of 4.3\,ppm~\cite{micke_coherent_2020}.

In this Letter, we present the $g$-factor measurement of the ground state of boron-like \hci{118}{Sn}{45+}, performed in the \alp{}~apparatus, achieving a precision of 0.5~parts-per-billion on a system with a much higher nuclear charge ($Z=50$).
\alp{} is a Penning-trap experiment, which is cooled to \SI{4.2}{\kelvin} by liquid helium.
The apparatus is placed inside a warm-bore 4-Tesla superconducting magnet.
Optical access is integrated from below and is used for irradiation of microwaves.
From above, the Penning trap is connected to a room-temperature beamline, allowing for ion injection from external ion sources.
In order to maintain ultra-high vacuum in the measurement trap, a cryogenic valve is utilized to seal the trap region after ion injection, further details can be found in Ref.~\cite{sturm_alphatrap_2019}.
Although the influx of gas from the room-temperature section is greatly reduced, the HCI still occasionally collide with the remaining neutral background gas, leading to a reduction of their charge state through charge exchange.
The boron-like \hci{118}{Sn}{45+} ion used for the presented measurements is the product of a double electron capture from a previously stored lithium-like ion.
Based on the observed electron capture events with highly charged tin (charge-state $q>45$), a mean storage time of $41^{+20}_{-12}\,\mathrm{days}$ has been obtained.
To first order, the electron-capture cross section, and hence the storage time, follow a linear behavior with $q$~\cite{kravis_single-_1995}.
Therefore, under these conditions even the heaviest ions in very high charge states, e.g. \hci{}{Pb}{81+}, will have storage times of around 20 days, enough to allow high-precision spectroscopy in the setup.

In the Penning trap, the charged particle undergoes oscillations in a superposition of three independent harmonic motions.
These come from the interplay of the strong magnetic field along with a quadrupolar electric field.
In the electric field, the particle oscillates with frequency $\nu_z$, with the $z$ axis defined by the direction of the magnetic field.
The interaction between electric and magnetic field can be described by two circular frequencies called modified cyclotron frequency $\nu_+$ and magnetron frequency $\nu_-$.
The three eigenfrequencies can be related to the free-space cyclotron frequency $\nu_\mathrm{c}$, as shown by the invariance theorem $\nu_\mathrm c^2 = \nu_+^2 +\nu_z^2+\nu_-^2$~\cite{brown_geonium_1986}.
The free-space cyclotron frequency establishes a connection to the magnetic field via the charge-to-mass ratio $q/m_\mathrm{ion}$ and the strength of the magnetic field $\nu_\mathrm c = q B_0/(2 \pi m_\mathrm{ion})$.
Furthermore, the ground-state energy level of a particle with angular momentum will split in the magnetic field.
The energy separation is proportional to its Larmor frequency, which is proportional to the magnetic field and the $g$ factor $\nu_\mathrm L = g B_0 e /(4 \pi m_\mathrm e)$, where $e$ and $m_e$ represent the charge and mass of the electron, respectively.
Combining the equations for the Larmor and the free-space cyclotron frequency eliminates the magnetic field term, and consequently the $g$ factor can be experimentally obtained from
\begin{equation}
    g = 2 \frac{\nu_\mathrm L}{\nu_\mathrm c} \frac q e \frac{m_\mathrm e}{m_\mathrm{ion}}= 2 \Gamma_0 \frac q e \frac{m_\mathrm e}{m_\mathrm{ion}}.\label{eq:g}
\end{equation}
In this expression, the ratio $\Gamma_0 = \frac{\nu_\mathrm L}{\nu_\mathrm c}$ has to be measured while the other values are obtained from literature or prior measurements~\cite{tiesinga_codata_2021,kramida_nist_2021,morgner_stringent_2023}. 

The measurement has been performed in a double-trap setup, where in the so-called analysis trap (AT) the spin state is measured non-destructively, while in the other one, called precision trap (PT), $\Gamma_0$ is measured with high precision by probing various frequency ratios $\Gamma$ to determine the center of the resonance.
A cut model of the trap stack is depicted in Fig.~\ref{fig:trap}\textbf{a}.
The PT is a cylindrical 18\,mm diameter 7-electrode trap, the AT positioned below is a 6\,mm 5-electrode design.
\begin{figure}[t]
    \centering
    \includegraphics{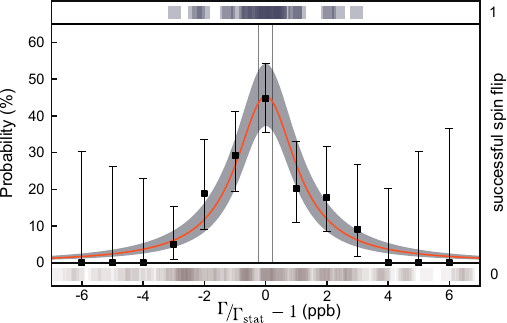}
    \caption{
    Spin-flip resonance of \hci{118}{Sn}{45+}.
    Individual measurement points are represented by the squares at the top and bottom for successful and unsuccessful attempts, respectively.
    To guide the eye, a binned set of the data with binomial error bars is shown.
    The data is fitted with a Lorentzian curve using a maximum-likelihood method.
    The vertical lines show the 1-$\sigma$ confidence interval of the center.
    }
    \label{fig:res}
\end{figure}
The electric and magnetic field in the PT is characterized with the ion by analyzing the frequency shifts for different modified cyclotron and magnetron radii~\cite{ketter_first-order_2014}.
The dominant anharmonicity coefficients of the electric field series expansion are evaluated to $C_4<1.2\cdot 10^{-4}$ and $C_6<1.3\cdot 10^{-2}$.
Furthermore, the magnetic field exhibits high homogeneity with a first-order magnetic field gradient of $B_1 = \SI[per-mode=symbol]{2.64(3)}{\milli\tesla\per\meter}$ and a second-order inhomogeneity $B_2 < \SI[per-mode=symbol]{10}{\milli\tesla\per\square\meter}$.
These properties ensure negligible systematic uncertainties on $\Gamma_0$ ($<$1\,ppt), as described in Ref.~\cite{morgner_stringent_2023}.
The particle detection and the motional frequency measurements in either trap is done via a superconducting resonator circuit, which amplifies the induced image currents and provides an ion signal at axial frequencies.
In resonance, the particle thermalizes with the circuit, and in the Fourier spectrum a \textit{dip} feature appears (see Fig.~\ref{fig:trap}b), which can be fitted to measure the axial frequency of the ion.
To detect the radial modes, they are coupled to the axial motion with a radio frequency field, allowing readout via the axial resonator~\cite{wineland_principles_1975}.
A phase-sensitive technique is employed to achieve greater precision on the modified cyclotron frequency~\cite{sturm_phase-sensitive_2011}.
In the PT, the motional frequencies are around $\nu_+\approx\SI{23}{\mega\hertz}$, $\nu_-\approx\SI{9}{\kilo\hertz}$, $\nu_\mathrm z\approx\SI{652}{\kilo\hertz}$.

In the AT, the center ring electrode, made of iron-cobalt, produces a magnetic bottle with a strong second-order magnetic field coefficient $B_2 \approx \SI[per-mode=symbol]{43}{\kilo\tesla\per\square\meter}$, which is used to measure the spin state of the ion.
Due to the continuous Stern-Gerlach effect~\cite{dehmelt_continuous_1986}, the interaction between the inhomogeneous magnetic field and the magnetic moment of the ion creates an additional force.
Consequently the axial frequency shifts depending on the spin orientation.
The frequency change $\Delta \nu_z$ for a change of the spin orientation $\Delta m_j = \pm 1$ is given by
\begin{equation}\label{eq:deltawz}
    \Delta \nu_z \approx \frac {B_2\hbar\,e}{8 \pi^2 m_\mathrm{ion} \nu_z m_e} g \Delta m_j.
\end{equation}
It is proportional to $g$ and the inverse mass, rendering it more challenging for heavier systems and those with a smaller $g$ factor such as boron-like systems.
Consequently, in boron-like tin, the frequency shift is only \SI{105}{\milli\hertz}.
To discern this minute change, we employ a phase-sensitive measurement technique for axial frequencies~\cite{stahl_phase-sensitive_2005}. 
A microwave at the Larmor frequency of the ions ($\approx\SI{35}{\giga\hertz}$) is used to drive the transition, thereby changing the spin state.
\begin{table}[b]
    \caption{Experimental error budget and systematic shifts (in ppt) are shown. Further effects are smaller than 1\,ppt.}
    \centering
    \begin{tabularx}{\columnwidth}{l l l}
     \hline\hline
        \textbf{Parameter} & \hspace{0.5cm} \textbf{Rel. shift}  &  \hspace{0.5cm} \textbf{Unc.} \\ 
        \hline
        $\Gamma_0 = \nu_\text{L}/\nu_\text{c}$ error budget: & &\\ 
        \hspace{0.5cm}$\nu_-$ measurement &\hspace{0.5cm}  - & \hspace{0.5cm} \phantom{0}\phantom{0}\SI{4.9}{}\\
        \hspace{0.5cm}Relativistic shift \cite{ketter_classical_2014} & \hspace{0.5cm} \phantom{0}\SI{20.0}{} & \hspace{0.5cm} \phantom{0}\phantom{0}\SI{4.0}{}\\
        \hspace{0.5cm}Image-charge shift \cite{schuh_image_2019} & \hspace{0.5cm} \SI{148}{} & \hspace{0.5cm} \phantom{0}\phantom{0}\SI{7.5}{} \\ 
        \hspace{0.5cm}$\nu_z$ line shape & \hspace{0.5cm} - & \hspace{0.5cm} \phantom{0}\SI{22}{}\\ 
        \hspace{0.5cm}statistical uncertainty & \hspace{0.5cm} -  & \hspace{0.5cm} \SI{240}{}\\ \hline
        $g$-factor error budget: & \hspace{0.5cm} &\\ 
        \hspace{0.5cm}Total $\Gamma_0$ uncertainty & \hspace{0.5cm} &\hspace{0.5cm} \SI{241}{}\\ 
        \hspace{0.5cm}Electron mass \cite{tiesinga_codata_2021} & \hspace{0.5cm}  & \hspace{0.5cm} \phantom{0}\SI{29}{} \\ 
        \hspace{0.5cm}\textsuperscript{118}Sn\textsuperscript{45+} mass~\cite{morgner_stringent_2023,kramida_nist_2021,tiesinga_codata_2021}& \hspace{0.5cm} & \hspace{0.5cm} \SI{475}{}\\   \hline \hline 
    \end{tabularx}
    \label{tab:ExpData}
\end{table}
The axial phases are measured both before and after the microwave irradiation to see a phase jump due to the flip of the spin state.
With the observed frequency stability of \SI{13}{\milli\hertz}, the spin state is determined in the AT with a 100\,\% fidelity throughout the whole measurement campaign.
More importantly, this shows that even in heavier (boron-like) systems, e.g. boron-like lead with a \SI{50}{\milli\hertz} frequency change, the spin state can be detected with high fidelity.
Further details are given in Fig.~\ref{fig:trap}.

A measurement cycle to determine $\Gamma_0$ starts in the AT where the spin state of the ion is determined.
Afterwards, the ion is adiabatically transported into the PT where we precisely measure the eigenfrequencies of its motional modes to determine $\nu_\mathrm c$.
Simultaneously, a microwave is irradiated at a random frequency near the expected Larmor frequency ($\approx\SI{36.3}{\giga\hertz}$ in the PT), which can induce a spin flip.
Subsequently in the AT, it is tested whether the microwave in the PT changed the spin state.
This cycle is repeated multiple times, probing different frequency ratios $\Gamma = \frac{\nu_{\mathrm{MW}}}{\nu_\mathrm c}$. 
For the presented measurement, this sequence was repeated in total 195 times of which 35 PT microwave attempts successfully induced a spin flip.
The resulting resonance is shown in Fig.~\ref{fig:res}.
A maximum-likelihood analysis is used to determine the free parameters: the width, the center position and the amplitude.
Fundamentally, the resonance follows a convolution of a Gaussian and a Lorentzian lineshape.
The Gaussian component arises from the magnetic field jitter and is known from previous measurements taken in similar conditions (relative full width at half maximum (FWHM) $\hat \approx~ 5.6\cdot 10^{-10}$, see Ref.~\cite{morgner_stringent_2023}).
The Lorentzian component is attributed to the two level system interacting with the applied microwave.
Since the natural linewidth of the transition is negligibly small, the width is given by the Rabi rate, and therefore by the electric field strength of the microwave.
As the observed FWHM (relative FWHM $\hat \approx~2.4\cdot 10^{-9}$) is significantly larger than the Gaussian component the resonance is fitted with a pure Lorentzian profile.
Further reduction of the Lorentzian linewidth could be achieved by reducing the microwave power, as has been realized in previous work~\cite{morgner_stringent_2023}.
But since the uncertainty of the mass is the biggest limitation on $g$ and there was no time to optimize the power due to problems with the insulation vacuum of the superconducting magnet, this was not implemented.
From the fit, the center is determined as $1\,539.242\,042\,61(37)$, which has to be corrected for systematic shifts.
The dominating ones are the image-charge shift, resulting from the image charges induced in the surrounding electrodes~\cite{schuh_image_2019}, and the relativistic correction where due to the velocity on a high cyclotron radius during the $\nu_+$ measurement the cyclotron frequency is slightly shifted~\cite{ketter_classical_2014}.

Other important systematic uncertainties, e.g. from the magnetron frequency uncertainty, or from the line shape in the axial frequency determination, are evaluated as explained in Ref.~\cite{morgner_stringent_2023} and are listed in Table~\ref{tab:ExpData}.

The final experimental value for the corrected $\Gamma_0$ ratio is:
    $\Gamma_0 = 1539.242\,042\,354(370)_\mathrm{stat}(37)_\mathrm{sys}$.
    Using Eq.~\eqref{eq:g}, and combining it with the masses reported in \cite{mohr_codata_2024,kramida_nist_2021,morgner_stringent_2023}, we derive the experimental $g$ factor as:
\begin{equation}
    g_\mathrm{exp} = 0.644\,703\,826\,493(155)_\mathrm{stat}(16)_\mathrm{sys}(307)_\mathrm{ext},
\end{equation}
with the brackets respectively being the statistical, systematic and external (uncertainty of the mass ratio $m_e/m_\mathrm{ion}$) 1-$\sigma$ confidence levels.
Note that higher-order Zeeman effects, which are large in light boron-like $g$ factors, become negligible in high-$Z$ elements as \hci{118}{Sn}{45+}~\cite{varentsova_interelectronic-interaction_2018}.

To compare with the experimental value, we calculated the $g$ factor of the $1s^2 2s^2 2p$ $^2 P_{1/2}$ ground state of the \Ion{118}{Sn}{45+} ion.
The one-photon exchange correction to the $g$ factor is calculated in a QED framework~\cite{Shabaev2002}, with a B-spline basis set with dual kinetic balance~\cite{Shabaev2002Dual,Cakir2020}.
Electron correlation effects are taken into account by means of a relativistic configuration interaction (CI) method employing a Dirac-Fock-Sturmian basis set as in Ref.~\cite{Shchepetnov2015,arapoglou_g_2019}.
The finite nuclear size (FNS) correction was calculated assuming the homogeneously charged sphere model of the nucleus, with the root-mean-square nuclear charge radius of $4.6519(21)$\,fm from Ref.~\cite{angeli_table_2013}.

The QED theory on the one-electron level is rather well developed, see e.g.~\cite{Pachucki2004,sapirstein_determination_2001,mohr_qed_1998,Sapirstein2023}.
The self-energy correction of the $2p_{1/2}$ valence electron, calculated first in the one-electron approximation and to all orders in $Z\alpha$, is the dominant QED contribution.
The effect of the screening on the SE of the valence electron was accounted for by means of an effective potential induced by the core electrons as in Ref.~\cite{Yerokhin2013}.
VP effects can typically be cast into two categories: a virtual $e^-e^+$ pair may correct the interaction of the bound electron with the nucleus (electric loop VP) or its interaction with the external magnetic field (magnetic loop VP).
The leading one-electron Uehling correction can be approximated as $-\frac{\alpha}{\pi}\frac{31}{840}(Z \alpha)^6$~\cite{Cakir2020}.
We calculate higher-order terms in $Z \alpha$ and the screening diagrams with the methods of~\cite{Cakir2020,Sikora2018,Lee2005,Dizer2023}.
Furthermore, the inclusion of two-loop QED contributes at the $(Z\alpha)^0$ (free-electron) level \cite{Brodsky1967,Grotch1973}, and a rigorous relativistic treatment of nuclear recoil~\cite{Glazov2018,Shchepetnov2015} is mandatory.
The recoil term is taken from Ref.~\cite{Glazov2018}.

\begin{table}[b]
  \caption{
    Contributions to the $g$~factor of boron-like \hci{118}{Sn}{45+}.
      The uncertainties given in parentheses indicate the uncertainty of the last digit(s). 
      If no uncertainty is given, all digits of the quoted value are significant.
      The error bars take into consideration both numerical and nuclear radius errors, in case of the one-photon exchange correction also the difference between Feynman and Coulomb gauge results. "pnt" stands for point nucleus.}
  \begin{ruledtabular}
    \begin{tabular}{ll}
      Contribution                           & \Center{Value} \\
      \hline                                 
      Dirac value                            & $\Pos{0.643\,484\,945\,97(1)}$ \\
      FNS              & $\Pos{0.000\,000\,055\,68(6)}$ \\
      Electron correlation:                  &                             \\
      \quad one-photon exchange $(1/Z)^{1}$, pnt & $\Pos{0.001\,937\,506\,837(1)}$ \\
      \quad one-photon exchange, FNS                         & $\Neg{0.000\,000\,068\,27(7)}$ \\
      \quad $(1/Z)^{2+}$, CI-DFS             & $\Neg{0.000\,011\,1(5)}$    \\
      Nuclear recoil                         & $\Neg{0.000\,003\,357(8)}$  \\
      One-loop QED:                          &                             \\
      \quad SE, $(1/Z)^{0}$                  & $\Neg{0.000\,700\,400}$     \\
      \quad SE, $(1/Z)^{1+}$                 & $\Neg{0.000\,005\,9(6)}$    \\
      \quad VP, $(1/Z)^{0}$:                 & $\Neg{0.000\,000\,136}$     \\
      \quad VP, $(1/Z)^{1+}$:                & $\Pos{0.000\,000\,175}$     \\
      Two-loop QED                           & $\Pos{0.000\,001\,18(6)}$   \\
      \hline
      Total theory                           & $\Pos{0.644\,702\,9(8)}$    \\
    \end{tabular}
  \end{ruledtabular}
  \label{tab:theory}
\end{table}

Finally, we arrive at a value for the boron-like $g$ factor from \textit{ab initio} theory:
\begin{equation}
    g_\mathrm{theo} = 0.644\,702\,9(8).
\end{equation}
The precision of the theoretical value is limited about equally by the many-photon exchange electron-electron interaction and the SE screening contributions.
Comparison with the measurement shows agreement at the 1.2-$\sigma$ level, showing that the theory is accurate, even in these much heavier systems.


In Refs.~\cite{shabaev_g-factor_2006,volotka_nuclear_2014} it was suggested that the fine-structure constant $\alpha$ can be determined employing a specific difference of the $g$ factors of the boron-like and hydrogen-like ion of the same isotope:
\begin{equation}
\label{eq:specdiff}
\Delta g = g^{1s^2 2s^2 2p_{1/2}} - \xi g^{1s} \,.
\end{equation}
The hydrogen-like value is scaled down with a prefactor~$\xi$ chosen to suppress the detrimental FNS contribution in the difference.
Thus, a theoretical uncertainty allowing a competitive determination of $\alpha$ can be reached.
The achievable relative uncertainty $\delta\alpha$ can be approximated as
\begin{equation}
    \delta \alpha \approx \frac{2}{(\alpha Z)^2} \sqrt{(\delta g_\mathrm{theo})^2+(\delta g_\mathrm{exp})^2},
\end{equation}
with $\delta g_\mathrm{exp/theo}$ as the relative uncertainty in the specific difference value of experiment and theory, respectively.

Considering only the experimental precision of 0.5\,ppb ($\delta g_\mathrm{theo}\rightarrow 0$), this would allow to extract $\alpha$ with a relative uncertainty of 7\,ppb, limited by the ${}^{118}$Sn mass, closely followed by the boron-like $\Gamma_0$ uncertainty.
For both quantities it has been shown that a measurement with an uncertainty below 30\,ppt is possible~\cite{sturm_high-precision_2014,heise_high-precision_2023}, which would allow the extraction of $\alpha$ with an uncertainty of 0.4\,ppb.
This could help to resolve the present disagreement seen with other methods~\cite{fan_measurement_2023}.
On the theoretical side, the evaluation of various effects needs to be improved, as Table~\ref{tab:theory} shows. Most urgently, two-photon exchange and QED screening diagrams, already well understood in Li-like ions~\cite{Yerokhin2020,Yerokhin2021,kosheleva_g_2022}, need to be computed in the boron-like case as well. The evaluation of two-loop QED corrections in a strong Coulomb field~\cite{Sikora2024} can be extended from the $1s$ to the $2p$ valence orbital in a straightforward manner. Further significant improvements beyond these terms are also necessary.

Interestingly, in boron-like tin, the tree-level and one-photon-exchange FNS effects show an almost 1-digit cancellation (second and fourth numerical entries in Table~\ref{tab:theory}, this effect is analogous to the cancellation of VP terms~\cite{Cakir2020}).
Such cancellation is much weaker in very heavy elements such as \hci{208}{Pb}{}, where a relatively large weighting factor $\xi$ is required, and hence a very precise calculation of the hydrogen-like $g$ factor.
For \hci{118}{Sn}{} we calculate $\xi_{\rm Sn}=-0.000\,86$, which is an order of magnitude smaller than for \hci{208}{Pb}{}~\cite{shabaev_g-factor_2006}.
Therefore, even without the specific difference, only considering boron-like tin, the FNS uncertainty is limiting a potential $\alpha$ determination to only 0.3\,ppb uncertainty.
This cancellation is even better in boron-like xenon, where $\xi_{\rm Xe}$ is on the order of $1\cdot 10^{-5}$. 
With such a small $\xi$, the specific difference is not necessary anymore, as the residual FNS uncertainty is already negligible in the boron-like value alone.
Although difficult, rigorous treatment of the QED effects therefore allows a theoretical prediction inherently insensitive to the FNS effects.
Thus, with elements near $Z=54$, a competitive determination of $\alpha$ appears to be more realistic than with very heavy ions.   


In summary, our measurement tests QED in the thus far unexplored regime of medium-to-high $Z$ boron-like ions.
The experiment is about a factor of 2000 more precise than the theoretical prediction, showing that it can be used as a benchmark for future advances in bound-state QED.
It marks a crucial data-point in the middle between light and very heavy boron-like systems.

The observed storage time, as well as the demonstrated high-fidelity spin-flip detection in the AT prove that $g$-factor measurements with hydrogen-like, lithium-like and even boron-like systems with highest $Z\geq82$ are feasible in the \alp~apparatus.
Such measurements could be used to test bound-state QED at highest $Z$.
Furthermore, the inherent cancellation of the tree-level and one-photon-exchange nuclear size terms brings an advantage for middle-$Z$ boron-like systems, optimally in Xe, making it a prime candidate for a potential $\alpha$ determination.
This cancellation also allows for more precise tests of the nuclear recoil effect through isotope shift measurements, as the finite nuclear size contribution was the primary source of uncertainty in previous hydrogen-like neon measurements~\cite{sailer_measurement_2022}.

This work was supported by the Max Planck Society (MPG), the International Max Planck Research School for Quantum Dynamics in Physics, Chemistry and Biology (IMPRS-QD), the German Research Foundation (DFG) Collaborative Research Centre SFB 1225 (ISOQUANT) and the Max Planck PTB RIKEN Center for Time, Constants, and Fundamental Symmetries.
This project has received funding from the European Research Council (ERC) under the European Union’s Horizon 2020 research and innovation programme under grant agreement number 832848 FunI. 
B. Tu would like to thank the Max Planck Partner Group project for their support. 
This work comprises parts of the PhD thesis work of J.M. submitted to Heidelberg University, Germany.
\bibliographystyle{apsrev4-1}

\end{document}